\documentclass{elsart}

\usepackage{graphicx}

\usepackage{amssymb,amsmath,citesort}

\newcommand{\beq}{\begin{equation}}
\newcommand{\eeq}{\end{equation}}
\newcommand{\bea}{\begin{eqnarray}}
\newcommand{\beas}{\begin{eqnarray*}}
\newcommand{\beau}[1]{\begin{equation} \label{#1} \begin{array}{rcl}}
\newcommand{\eea}{\end{eqnarray}}
\newcommand{\eeas}{\end{eqnarray*}}
\newcommand{\eeau}{\end{array} \end{equation}}
\newcommand{\bay}{\begin{array}}
\newcommand{\eay}{\end{array}}
\newcommand{\bals}{\begin{align*}}
\newcommand{\eals}{\end{align*}}

\newcommand{\vev}[1]{\langle #1 \rangle}

\begin{document}

\begin{frontmatter}


\title{Hadronization in cold nuclear matter\thanksref{talk}}
\thanks[talk]{Talk given at ``Hard Probes 2006'', Asilomar Conference
Grounds, Pacific Grove, CA (USA), June 9-16, 2006.}
\author{Alberto Accardi}
\address{Dept. of Physics and Astronomy, Iowa State U., Ames, IA
  50011, USA}

\begin{abstract}
I review a recently proposed scaling analysis of hadron suppression
in Deeply Inelastic Scattering on nuclear targets measured at the
HERMES experiment. The analysis can distinguish 2 competing
explanations for the observed suppression, namely, quark radiative
energy loss with long hadron formation times, and prehadron
nuclear absorption with hadronization starting inside the nucleus.
Experimental data are shown to favor short formation times and
prehadron absorption.
\end{abstract}

\begin{keyword}
Hadron formation time \sep radiative energy loss \sep prehadron absorption.
\PACS {25.30.-c}{} \sep {25.75.-q}{} \sep {24.85.+p}{} \sep {13.87.Fh}{} 
\end{keyword}
\end{frontmatter}


One of the most striking experimental discoveries in the heavy-ion
program at the Relativistic Heavy Ion Collider (RHIC) has been the
suppression of large transverse momentum hadron production in
nucleus-nucleus (A+A) collisions compared to proton-proton collisions
\cite{RHIC}. An analogous hadron suppression has been observed in
Deeply Inelastic Scattering on nuclear targets (nDIS)
\cite{EMC,HERMES1,HERMES2,JLAB}, where the
observable of interest is the hadron multiplicity ratio
\begin{align}
  R_M^h(z_h) = \frac{1}{N_A^{DIS}}\frac{dN_A^h(z_h)}{dz_h} \Bigg{/}
    \frac{1}{N_D^{DIS}}\frac{dN_D^h(z_h)}{dz_h} ,\
    \label{MultiplicityRatio}	   
\end{align}
i.e., the single hadron multiplicity on a target of mass number $A$ 
normalized to the multiplicity on a deuteron target as a function of
the hadron's fractional energy $z_h=E_h/\nu$, where $\nu$ is 
the virtual photon energy. 

On the theoretical side, 2 frameworks are presently competing
to explain the observed attenuation of hadron production in nDIS: 
quark energy loss, with hadron formation outside the nucleus
\cite{Wang,Arleo,Accardi:2005mm} and nuclear absorption, with
hadronization starting inside the nucleus
\cite{BG+AMP,AGMP05,Kopeliovich,Falteretal04}. 
Distinguishing between these 2 different pictures of the
space time evolution of hadronization is essential 
to correctly interpret hadron suppression in A+A collisions 
as due to parton-medium or hadron-medium interactions, and to
correctly extract properties of the produced Quark-Gluon Plasma (QGP)
such as its density or temperature from the measured hadron spectra. 
See Ref.~\cite{Accardi:2006ea} for a review. 

\begin{figure}
\centerline{
\parbox{13cm}{
  \vspace*{-0.1cm}
  \includegraphics[height=6.2cm,origin=c]{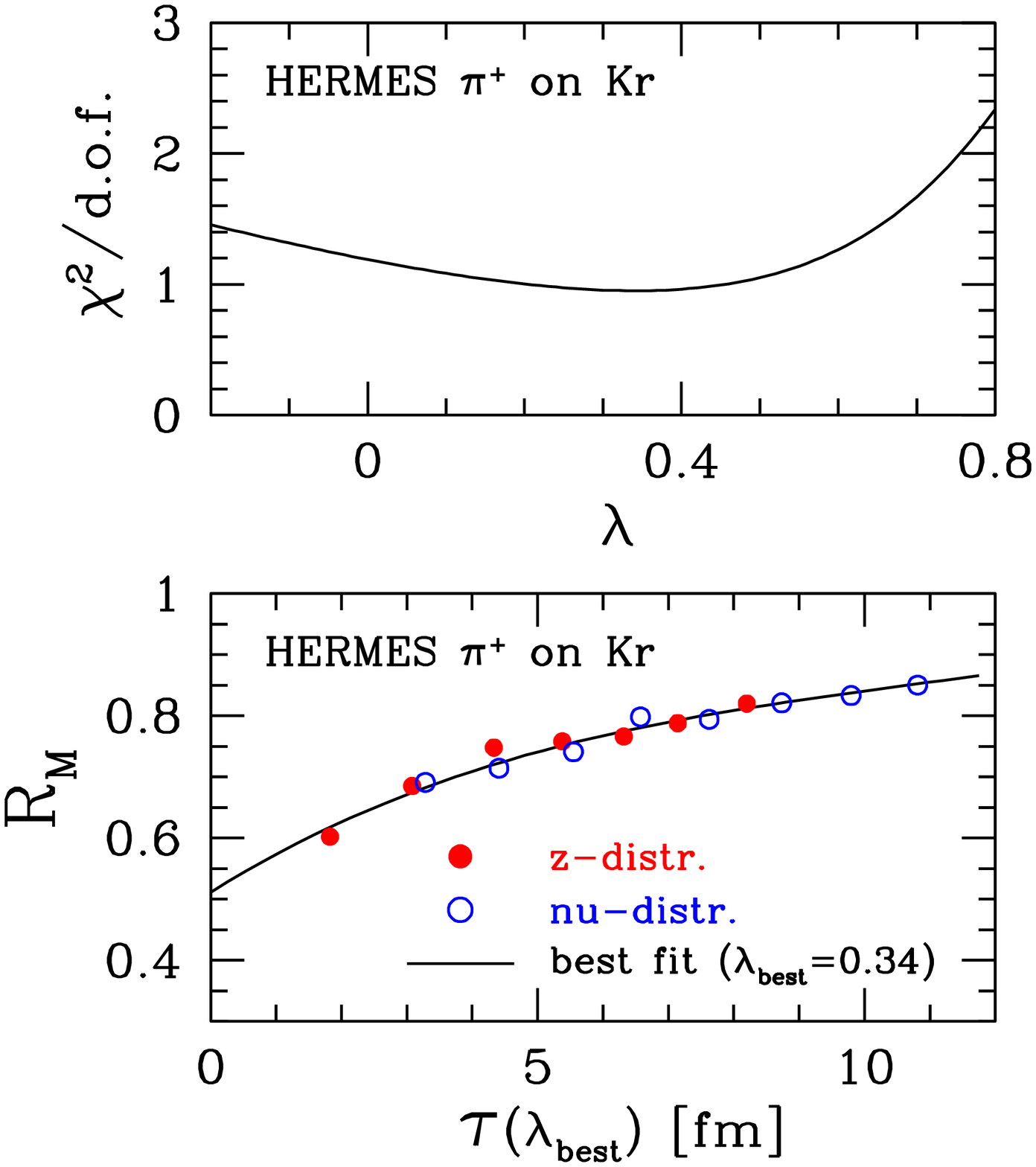}
  \hspace*{-.6cm}
  \includegraphics[height=6.2cm,origin=t]{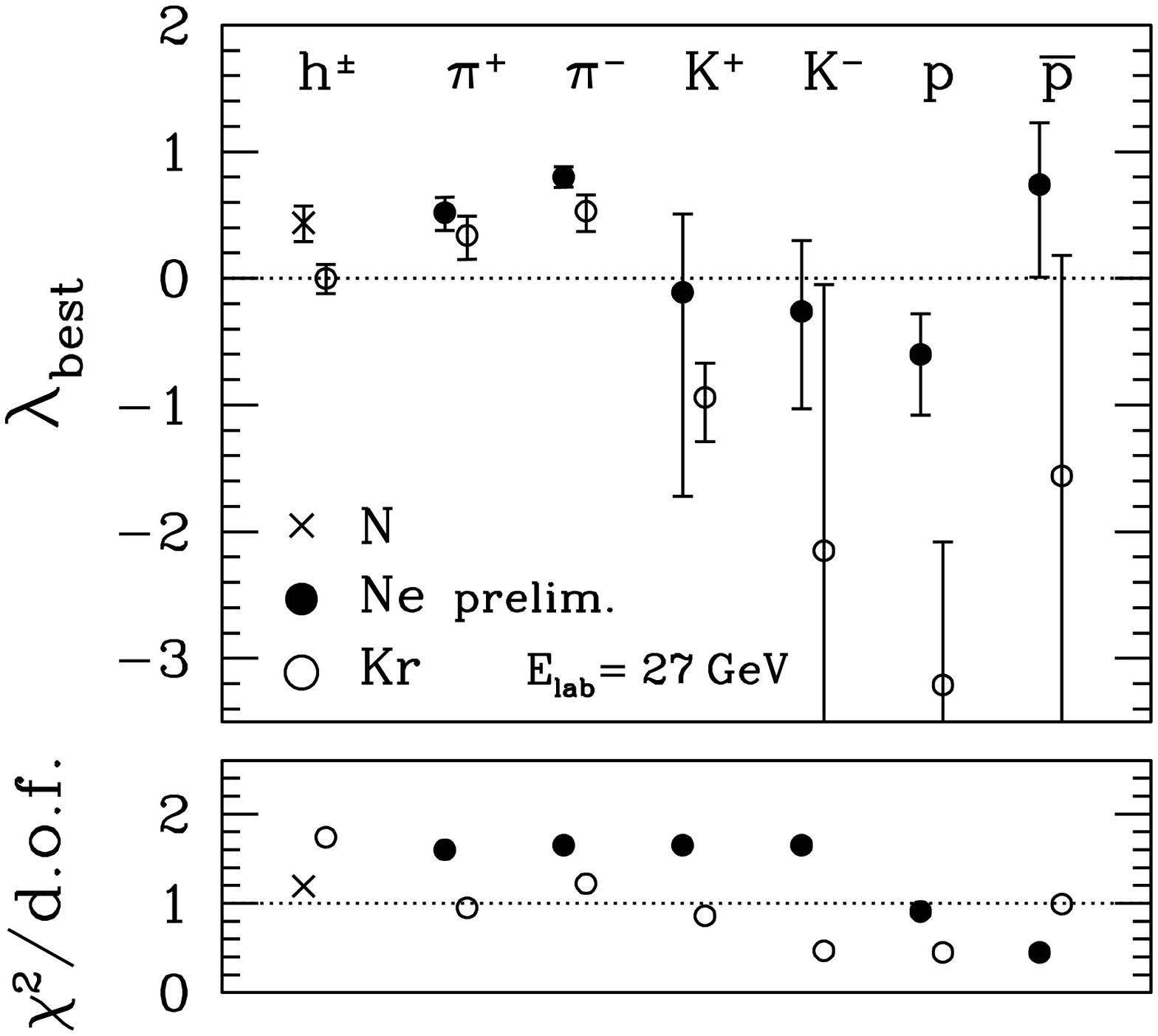} 
  }}
 \caption{Left: Extraction of $\lambda$ and scaling of $R_M$ HERMES
   data for $\pi^+$ on Kr \cite{HERMES1}. Right: Scaling exponents
   with 1$\sigma$ error bars, and $\chi^2$ per degree of freedom 
   extracted from HERMES data on charged and 
   identified hadrons at $E_{lab}=27$ GeV 
   \cite{HERMES1,HERMES2}. 
   }
 \label{fig:lambdafit} \label{fig:HERMESfit27}
\end{figure}

In Ref.~\cite{Accardi06}, I proposed a scaling analysis of the
experimental multiplicity ratio as a mean of distinguishing quark 
energy loss from nuclear absorption. Specifically, I conjecture that
$R_M$ should depend on $z_h$ and $\nu$ only as 
\begin{align}
  R_M = R_M[\tau] 
  \hspace*{0.5cm} \text{with} \hspace*{0.5cm}
  \tau = C\, z_h^\lambda (1-z_h) \nu \ .
 \label{eq:RMscaling}
\end{align}
The scaling exponent $\lambda$ is introduced as a way of
approximating and summarizing the scaling behavior of experimental
data and theoretical models. It can be obtained by a best fit
analysis of experimental data or theoretical computations, see
Fig.~\ref{fig:lambdafit}. 
The constant $C$ cannot be determined by the fit. 
A possible scaling of $R_M$ with $Q^2$ is not considered in the present
analysis.
As discussed below, the proposed functional
form of $\tau$ is flexible enough to
encompass both absorption models 
and energy loss models. The 2 classes of models are distinguished by 
the value of the scaling exponent: a positive $\lambda \gneqq 0$ is
characteristic of absorption models, while a negative $\lambda \lesssim
0$ is characteristic of energy loss models. Thus, the exponent
$\lambda$ extracted from experimental data can identify the leading
mechanism for hadron suppression in nDIS.

The scaling of $R_M$ is quite natural in the context of absorption models
\cite{BG+AMP,AGMP05,Kopeliovich,Falteretal04}.
In these models, hadronization is assumed to proceed in 2
steps. First, the struck quark neutralizes its color and becomes a
``prehadron'', with non-negligible inelastic cross-section with the
nuclear medium. Subsequently, and typically outside the nucleus, 
the prehadron collapses on the observed hadron wave function.
The nuclear absorption of the prehadron depends on the  
in-medium prehadron path length, which depends solely on
the prehadron formation time $\vev{t_*}$. In string models
\cite{BG+AMP,AGMP05}, as well as
in pQCD inspired computations \cite{Kopeliovich},
\begin{align}
  \vev{t_*} \propto f(z_h) (1-z_h) z_h \nu
\end{align}
which is well described by the proposed scaling variable $\tau$ with
$\lambda>0$. E.g., in the  Lund model $\lambda \approx 0.7$. 
In energy loss models \cite{Wang,Arleo,Accardi:2005mm}, the scaling is
less obvious. Hadronization is assumed to happen outside the
nucleus. Then hadron suppression is due to a reduction of the
available quark energy due to medium-induced gluon radiation. The
energy $\Delta E$ carried away by the radiated gluons is limited by
energy conservation to $\Delta E = (1-z_h)\nu$, which in turn implies
an approximate scaling 
of $R_M$ with $\tau = C (1-z_h) \nu$, i.e., with $\lambda \approx
0$. In practice it turns out that in energy loss models $\lambda
\lesssim 0$. See Ref.~\cite{Accardi06} for full details.
\begin{figure}[tb]
  \vspace*{0cm}
  \centerline{
  \includegraphics
    [height=6.4cm,bb=20 169 530 635,clip=true]
    {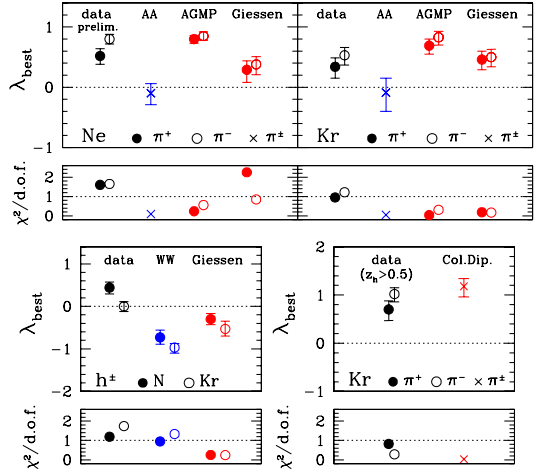}
  }
  \vspace*{0cm}
  \caption[]{
  Comparison of $\lambda$ from HERMES data
  \cite{HERMES1,HERMES2} and from theory models.  
  Energy loss models (blue points on line ): 
  AA \cite{Accardi:2005mm}, WW \cite{Wang}. Absorption
  models (red points): AGMP (pure absorption
  without $Q^2$-rescaling) \cite{AGMP05,Accardi:2005mm}, 
  Col.Dip. \cite{Kopeliovich}, Giessen \cite{Falteretal04}.
  \label{fig:datatheory}
 }
\end{figure}
The scaling exponents $\lambda_{\rm best}$ extracted from HERMES data
at $E_{lab}=27$ GeV \cite{HERMES1,HERMES2} for different
hadron flavors and nuclei  
are shown in Fig.~\ref{fig:HERMESfit27}. In all cases $\chi^2/{\rm
  d.o.f.} \lesssim 1.6$, which proves that $R_M$ scales with $\tau$.
The comparison of experimental and theoretical 
scaling exponents is shown in Fig.~\ref{fig:datatheory}.

In conclusion, experimental data on pion and 
charged hadron production have been shown to scale with $\tau$ and
exhibit $\lambda \gtrsim 0.4$. A discussed, this is a clear indication
of the dominance of the prehadron absorption
mechanism as opposed to the energy loss mechanism, or in other words
it is a signal of in-medium prehadron formation, with formation times
$\vev{t_*}\lesssim R_A$. The scaling variable $\tau$ can then be
interpreted as a measure of the formation time of the prehadron, the
color neutral precursor of the observed hadron. 
A more direct detection of in-medium hadronization, and a measurement
of the overall scale of the prehadron formation time, is possible by
looking at the hadron $p_T$-broadening, as proposed in
Ref.~\cite{Kopeliovich}. The outlined scaling analysis 
will be a useful cross-check of this measurement.
Establishing a scaling of the
prehadron formation time with $Q^2$, as predicted, e.g., in
Ref.~\cite{Kopeliovich}, will further constrain the hadronization
mechanism. A dedicated experimental analysis is needed to 
improve the reach and precision of the proposed scaling analysis.
Finally, note that the hadrons observed
at HERMES have energies $E_h = z_h \nu \approx 2-20$ GeV, which are
comparable to mid-rapidity hadrons at RHIC ($E_h\approx p_T$). Thus, at
RHIC one may expect hadronization to start inside the QGP.  

{\it Acknowledgments.} Work partially funded by the US
DOE grant no. DE-FG02-87ER40371. I am grateful to the
organizers for partial support.


\begin{thebibliography}{00}



\bibitem{RHIC} 
  I.~Arsene {\it et al.}  [BRAHMS],
  Nucl.\ Phys.\ A {\bf 757}, 1 (2005);
  B.~B.~Back {\it et al.} [PHOBOS],
  Nucl.\ Phys.\ A {\bf 757}, 28 (2005);
  J.~Adams {\it et al.}  [STAR],
  Nucl.\ Phys.\ A {\bf 757}, 102 (2005);
  K.~Adcox {\it et al.}  [PHENIX],
  Nucl.\ Phys.\ A {\bf 757} (2005) 184.

\bibitem{EMC}
  J.~Ashman {\it et al.}  [EMC],
  Z.\ Phys.\ C {\bf 52} (1991) 1.

\bibitem{HERMES1}
  A.~Airapetian {\it et al.}  [HERMES],
  Eur.\ Phys.\ J.\ C {\bf 20} (2001) 479 and
  Phys.\ Lett.\ B {\bf 577} (2003) 37. 

\bibitem{HERMES2}
  G.~Elbakian {\it et al} [HERMES], 
  Proceedings ``DIS 2003'', St.Petersburg, April 23-27, 2003; 
  V.T.~Kim and L.N.~Lipatov eds., page 597.  

\bibitem{JLAB}
  W.~K.~Brooks [CLAS], talk at Jefferson Laboratory Users Group Workshop,
  June 13, 2006.

\bibitem{Wang} 
  E.~Wang and X.~N.~Wang,
  Phys.\ Rev.\ Lett.\   {\bf 89} (2002) 162301.

\bibitem{Arleo}
  F.~Arleo,
  Eur.\ Phys.\ J.\ C {\bf 30} (2003) 213 
  and
  JHEP {\bf 0211} (2002) 044.

\bibitem{Accardi:2005mm}
  A.~Accardi, 
  to appear in Acta Phys. Hung.  [arXiv:nucl-th/0510090].

\bibitem{BG+AMP}
  A.~Bialas and M.~Gyulassy,
  Nucl.\ Phys.\ B {\bf 291} (1987) 793; 
  A.~Accardi, V.~Muccifora and H.~J.~Pirner,
  Nucl.\ Phys.\ A {\bf 720}, 131 (2003).

\bibitem{AGMP05}
  A.~Accardi, D.~Grunewald, V.~Muccifora and H.~J.~Pirner,
  Nucl.\ Phys.\ A {\bf 761} (2005) 67

\bibitem{Kopeliovich}
  B.~Z.~Kopeliovich, J.~Nemchik, E.~Predazzi and A.~Hayashigaki,
  Nucl.\ Phys.\ A {\bf 740} (2004) 211; 
  B.~Z.~Kopeliovich, J.~Nemchik and I.~Schmidt,
  arXiv:hep-ph/0608044.

\bibitem{Falteretal04}
  T.~Falter, W.~Cassing, K.~Gallmeister and U.~Mosel,
  Phys.\ Rev.\ C {\bf 70} (2004) 054609.

\bibitem{Accardi:2006ea}
  A.~Accardi, 
  talk at Hot Quark 2006, Villasimius, Sardegna (Italy),
  May 15-20, 2006, to appear in Eur. Phys. J. C  
  [arXiv:nucl-th/0609010].

\bibitem{Accardi06}
  A.~Accardi,
  arXiv:nucl-th/0604041.

\end{thebibliography}
\end{document}